\documentclass[times, review, 10pt]{elsarticle}

\usepackage{amssymb}
\usepackage{amsmath}

\usepackage[pagebackref,breaklinks,colorlinks,citecolor=eccvblue]{hyperref}
\usepackage{multirow} 
\def\eg{{\it{e.g.}}}
\def\etal{{\it{et al.}}}

\usepackage{subcaption}

\journal{Pattern Recognition}

\begin{document}

\begin{frontmatter}



\title{MMoFusion: Multi-modal Co-Speech Motion Generation with Diffusion Model}

\author[1,2]{Sen Wang} 
\ead{52285901013@stu.ecnu.edu.cn}
\author[3]{Jiangning Zhang}
\author[1]{Xin Tan}
\author[2]{Zhifeng Xie}
\author[3,4]{Chengjie Wang}
\author[1,4]{Lizhuang Ma}


\affiliation[1]{organization={School of Computer Science and Technology, East China Normal University},
            city={Shanghai},
            country={China}}
\affiliation[2]{organization={Shanghai University},
            city={Shanghai},
            country={China}}
\affiliation[3]{organization={Tencent Youtu Lab},
            country={China}}
\affiliation[4]{organization={Department of Computer Science and Engineering, Shanghai Jiao Tong University},
            city={Shanghai},
            country={China}}

\fntext[]{This work was done when Sen Wang interned at Tencent Youtu Lab.}

\begin{abstract}
The body movements accompanying speech aid speakers in expressing their ideas.
Co-speech motion generation is one of the important approaches for synthesizing realistic avatars. 
Due to the intricate correspondence between speech and motion, generating realistic and diverse motion is a challenging task. 
In this paper, we propose \textbf{MMoFusion}, a \textbf{M}ulti-modal co-speech \textbf{Mo}tion generation framework based on dif\textbf{Fusion} model to ensure both the authenticity and diversity of generated motion.
We propose a progressive fusion strategy to enhance the interaction of inter-modal and intra-modal, efficiently integrating multi-modal information. 
Specifically, we employ a masked style matrix based on emotion and identity information to control the generation of different motion styles. 
Temporal modeling of speech and motion is partitioned into style-guided specific feature encoding and shared feature encoding, aiming to learn both inter-modal and intra-modal features.
Besides, we propose a geometric loss to enforce the joints' velocity and acceleration coherence among frames.
Our framework generates vivid, diverse, and style-controllable motion of arbitrary length through inputting speech and editing identity and emotion. 
Extensive experiments demonstrate that our method outperforms current co-speech motion generation methods including upper body and challenging full body. Our code and model will be released at  \href{https://mmofusion.github.io/}{this URL}.
\end{abstract}


\end{frontmatter}



\section{Introduction}
\label{sec:intro}
\begin{figure*}[h]
\centering
\includegraphics[width=0.98\textwidth]{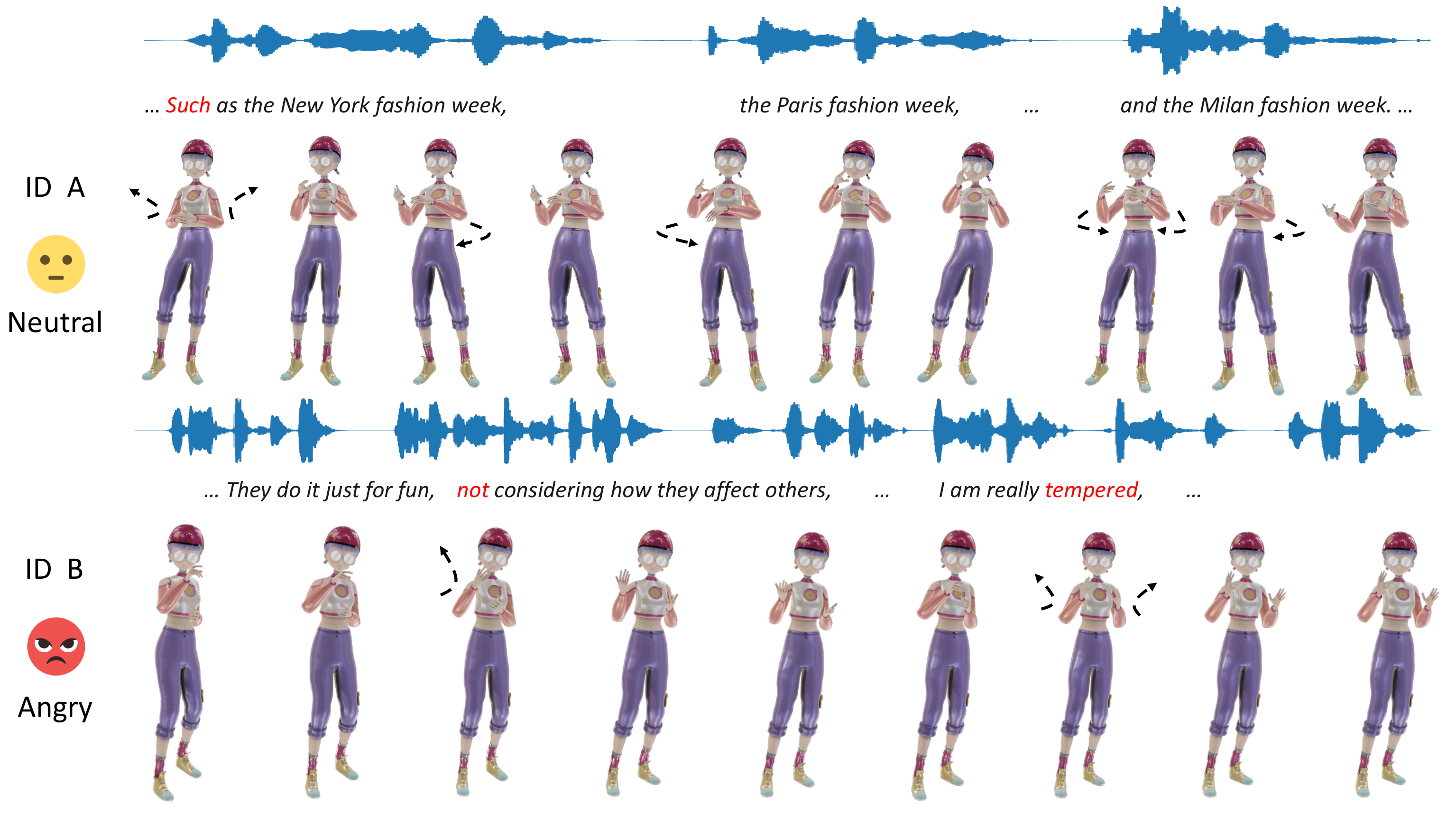}
\caption{
Our MMoFusion framework generates realistic, coherent, and diverse motions conditioned on speech, editable identities, and emotions. The top and bottom show motion results with different identities and emotions.
}
\label{fig:intro_1}
\end{figure*}

It is common for individuals to complement their speech with bodily movements, enhancing their ability to convey thoughts~\cite{cassell1999speech, kucherenko2021large}. 
Co-speech motion generation aims to synthesize realistic virtual avatars and can be utilized in entertainment, education, and social interaction.
The key to co-speech motion generation lies in the realism and diversity of the generated motion. 
To model the complex correspondence between speech and motion, various methods based on Generative Adversarial Networks (GANs)~\cite{habibie2021learning, yoon2020speech} and Variational Autoencoders (VAEs)~\cite{li2023audio2gestures, yi2023generating} have been proposed. 
However, these methods often constrain the learned distribution, limiting their ability to generate diverse motion. 
Contrastingly, diffusion models are not bound by assumptions about the target distribution, making them well-suited for modeling the many-to-many distribution matching problem in co-speech motion generation~\cite{tevet2023human}. 

\begin{figure}[t]
\centering
\includegraphics[width=0.96\linewidth]{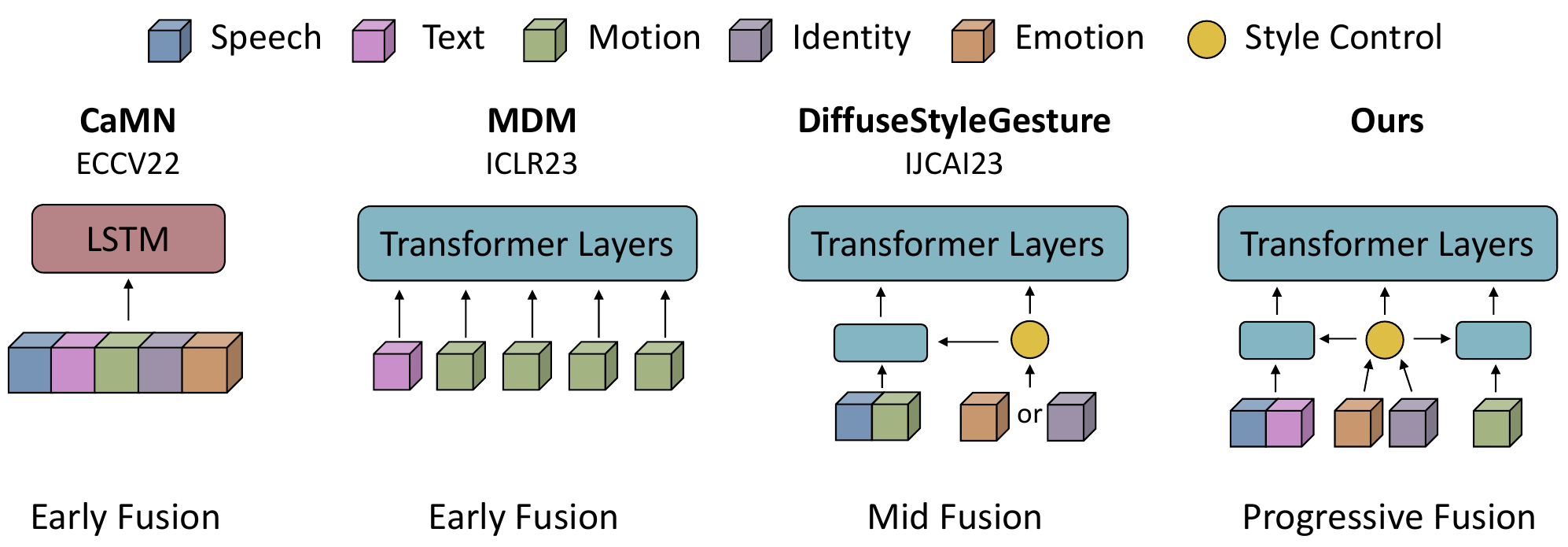}
\caption{
Comparison of our method with existing multi-modal motion generation methods. 
\textbf{Early Fusion}: CaMN\cite{liu2022beat} uses simple concatenation, MDM\cite{tevet2023human} utilizes conditional token. 
\textbf{Mid Fusion}: DiffuseStyleGesture~\cite{yang2023diffusestylegesture} leverages cross-local attention to establish intermediate representations.
We propose a \textbf{Progressive Fusion Strategy} to fully learn multi-modal features.
}
\label{fig:intro_2}
  \vspace{-1.5em}
\end{figure}

Harnessing diffusion models to generate convincingly realistic motion is a challenging task~\cite{tevet2023human,zhu2023taming}. 
An intuitive solution is to introduce additional relevant multi-modal information to produce high-fidelity motion such as transcript, identity, and emotions. 
The correspondence between speech and motion is explicitly decoupled using different modal information, such as differences in body motion between individuals and emotions in Figure \ref{fig:intro_1}.
As shown in Figure \ref{fig:intro_2}, multi-modal fusion approaches constitute one of the crucial factors influencing motion generation. 
\cite{liu2022beat} utilizes a cascade motion network (CaMN) to concatenate multi-modal features.
Motion Diffusion Model (MDM)~\cite{tevet2023human} utilizes conditional tokens to guide the diffusion model to generate motion.
As described in~\cite{nagrani2021attention}, these ``early fusion'' methods may introduce redundant information. 
``Mid fusion'', on the other hand, is conducive to extracting more significant mapping relationships. 
DiffuseStyleGesture (DSG)~\cite{yang2023diffusestylegesture} utilizes cross-local attention to establish intermediate representations and employ style control to guide motion generation, but we argue that the intermediate representation of speech and motion still harbors redundancy in the context of multi-modal fusion.
As it does not consider the specific information in speech and motion, respectively.

In this paper, to establish a more efficient ``mid fusion'', we propose a Progressive Fusion Strategy (PFS). 
It encompasses specific feature encoding and shared feature encoding between speech and motion features.
For the former, speech and motion features are separately encoded to extract specific information, reducing the impact of unnecessary fine-grained features and high-frequency noise. 
For the latter, we employ cross-attention\cite{vaswani2017attention} to extract shared features and aggregate specific and shared features to obtain hybrid features to generate motion.
Secondly, to enhance style control, we employ a masked style matrix calculated based on emotion and identity information to guide the two stages of feature fusion. 
This approach explicitly provides style cues during the fusion process and also performs classifier-free guidance\cite{ho2021classifier} to increase the motion diversity.
Additionally, to generate smoother motion, we propose a geometric loss that includes joint velocity and acceleration. 
Finally, we design a long sequence sampling to reduce inference time and generate consistent motion of arbitrary length. 

In summary, our main contributions are: 
\begin{itemize}
\item We propose a Multi-modal Co-speech Motion Generation Framework based on the Diffusion Model (MMoFusion), which employs a Progressive Fusion Strategy (PFS) including specific feature encoding and shared feature encoding to refine and fully learn multi-modal information.
\item In the process of PFS, we utilize a masked style matrix to guide multi-modal fusion and further control motion styles.
To generate coherent motion, we propose a geometric loss that includes joint velocity and acceleration. 
\item Extensive experiments demonstrate that our framework can generate vivid, diverse, and style-controllable motion that outperforms existing co-speech motion generation methods including upper body and full body.
\end{itemize}

\section{Related Work}
\label{sec:related_work}
\subsection{Human Motion Generation}

Human motion generation can be guided by various conditions such as text~\cite{guo2022generating, guo2022tm2t, petrovich2022temos, zhang2023t2m}, actions~\cite{guo2020action2motion, petrovich2021action}, and audio~\cite{ginosar2019learning, liu2022learning, li2021ai}. 
However, the non-deterministic mapping between audio and gestures poses a significant challenge. 
To this end, our primary focus lies on \textbf{co-speech motion generation}.
Motion matching \cite{clavet2016motion} is widely used for generating gestures ~\cite{habibie2022motion, yang2023qpgesture, ao2022rhythmic}.
However, these methods typically rely on time-consuming motion matching library.
Some RNN-based methods ~\cite{ferstl2018investigating, yoon2019robots, liu2022learning, liu2022beat} have been proposed, but they suffer from error accumulation. 
Other approaches leverage GANs ~\cite{habibie2021learning, yoon2020speech}, VAEs ~\cite{li2023audio2gestures, yi2023generating, gu2024orientation}, Transformer~\cite{dai2023kd} to produce natural motion. 
For instance, Yoon \etal{} \cite{yoon2020speech} train a adversarial network by using multi-modal information to generate human gestures. 
Li \etal{} \cite{li2023audio2gestures} divide the latent space into shared code and motion-specific code to train a generator.

\textbf{Motion Diffusion Models.}
Diffusion probability model has achieved significant results in both unconditional and conditional image generation ~\cite{ho2020denoising, ho2021classifier, rombach2022high, ruiz2023dreambooth}. 
For motion generation, \cite{zhang2022motiondiffuse}, \cite{chen2023executing} and \cite{dabral2023mofusion} introduce text-conditioned diffusion models, while \cite{zhu2023taming} proposes a audio-conditioned diffusion model. 
Moreover, several methods \cite{tseng2023edge, ao2023gesturediffuclip, yang2023diffusestylegesture, yin2023emog, fu2024mambagesture, mao2024mdt} leverage multi-modal information as conditions to generate high-quality motion sequences.
Specifically, 
GestureDiffuCLIP \cite{ao2023gesturediffuclip} takes text, motion, and video inputs as style prompts, and incorporates them via an adaptive instance normalization (AdaIN) layer \cite{huang2017arbitrary}.
EMoG \cite{yin2023emog} utilizes audio to extract emotional feature and a joint correlation-aware transformer to generate motion.
However, current methods rely solely on operations like feature concatenation \cite{zhu2023taming, yang2023diffusestylegesture}, token guidance \cite{tevet2022human}, or cross-attention \cite{ao2023gesturediffuclip} to integrate multi-modal features. 
In contrast to above methods, we introduce a multi-modal progressive fusion strategy for generating more realistic and vivid human motion.

\subsection{Multi-Modal Learning}

Multi-modal learning is related to many scenarios, including including face generation~\cite{xu2023high, zhang2021real, wu2023audio}, and human motion generation~\cite{tseng2023edge, ao2023gesturediffuclip, yang2023diffusestylegesture, yin2023emog}.
Zhang \etal{} \cite{zhang2019neural} use cross-attention to leverage visual and text features for machine translation. 
Lu \etal{} \cite{lu2020cross} introduce specific and shared feature transformation algorithms for person re-identification. 
Qin \etal{} \cite{qin2020feature} similarly used feature projection to learn specific and shared features for text classification. 
Xie \etal{} \cite{xie2023boosting} employ learnable frequency features to guide spatial features for nighttime scene segmentation. 
Recently, Ruan \etal{} \cite{Ruan_2023_CVPR} propose a multi-modal diffusion model for video and audio generation.
Multi-modal fusion is a crucial technique in multi-modal learning \cite{8269806}. 
To merge multi-modal features, 
several methods have been proposed that fall into the category of "early fusion" ~\cite{zadeh2017tensor, liu2018efficient, hou2019deep}.
Zadeh \etal{} \cite{zadeh2017tensor} propose a matrix-based multi-modal tensor fusion network. 
Liu \etal{} \cite{liu2018efficient} introduce low-rank multi-modal fusion to reduce the model parameters. 
Hou \etal{} \cite{hou2019deep} introduce the Polynomial Tensor Pooling (PTP) block, which utilizes high-order tensors to integrate multi-modal features. 
Additionally, there are "late fusion" methods that represent different modalities as low-dimensional semantic vectors and compute their semantic distances ~\cite{huang2013learning, shen2014learning}.
Nagrani \etal{} \cite{nagrani2021attention} firstly introduce "mid fusion" and shared tokens to enhance multi-modal fusion performance. 
Inspired by the above methods ~\cite{zadeh2017tensor, nagrani2021attention, lu2020cross, Ruan_2023_CVPR}, we introduce multi-modal fusion into human motion generation and diffusion model.

\section{Method} \label{sec:method}

\begin{figure*}[!t]
\centering
\includegraphics[width=0.96\textwidth]{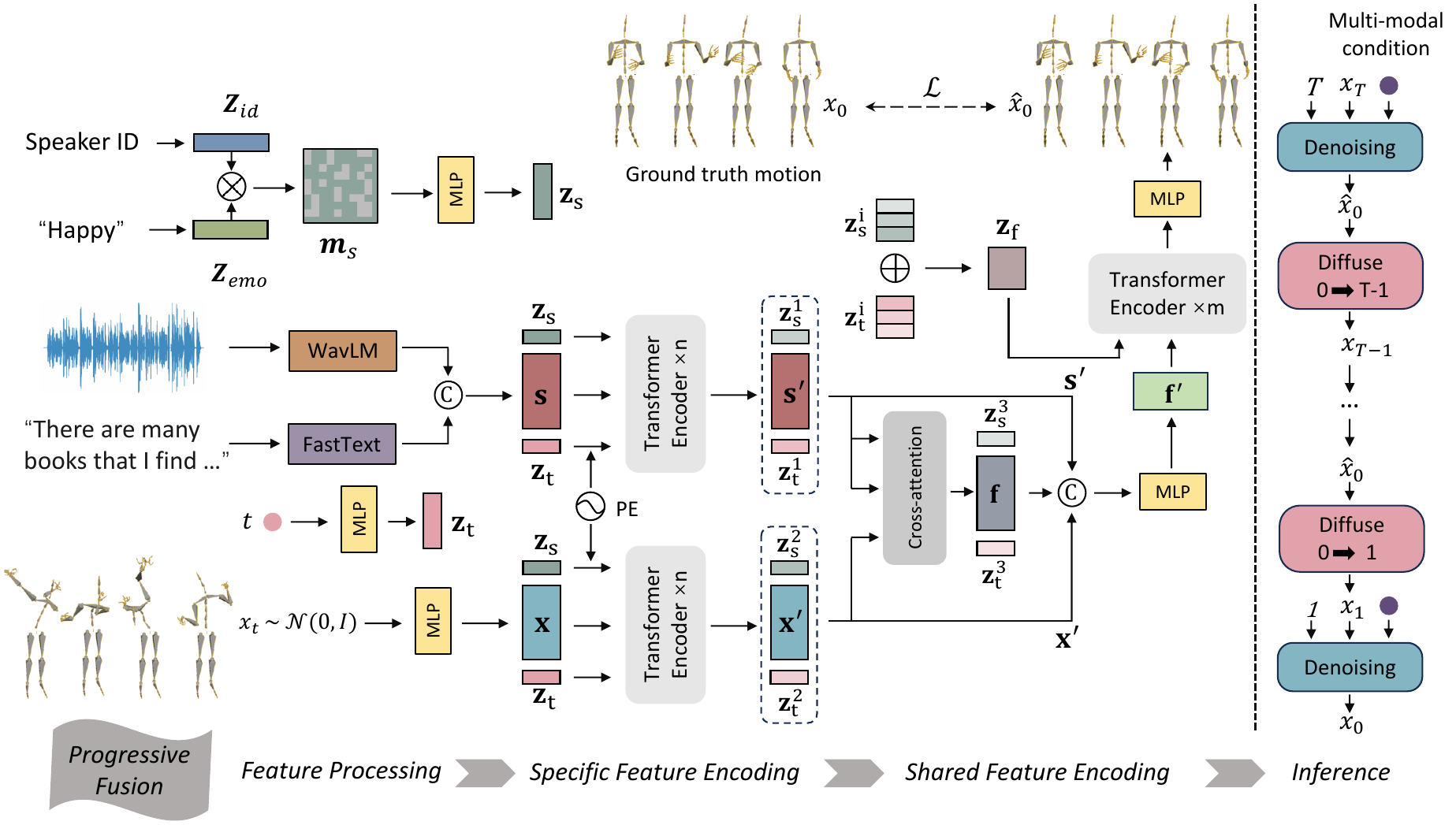}
\caption{\textbf{Overview of MMoFusion framework.}
We propose a Progressive Fusion Strategy (PFS) to fuse multi-modal information including
\textbf{\textit{1) Feature Processing.}}
A noisy motion sequence $x_t$ at time step $t$ is fed into the diffusion model conditioning on multi-modal information. 
Speech feature $\mathbf{s}$ is obtained by concatenating the transcript and audio features extracted from pre-trained models.
We utilize a masked style matrix $\mathbf{m}_{s}$ to guide motion generation.
It is mapped into a style token $\mathbf{z}_{s}$ during the whole multi-modal fusion.
\textbf{\textit{2) Specific Feature Encoding.}}
Speech feature $\mathbf{s}$ and motion feature $\mathbf{x}$ are encoded, respectively to obtain the specific features $\mathbf{s}'$ and $\mathbf{x}'$.
\textbf{\textit{3) Shared Feature Encoding.}}
Shared feature $\mathbf{f}$ is obtained by fusing specific features with cross-attention. 
Finally, the motion $\hat{x}_{0}$ is generated by the hybrid feature $\mathbf{f}'$ aggregated from the specific and shared features and guided by three different style tokens $\mathbf{z}_{s}^{i}$ and time tokens $\mathbf{z}_{t}^{i}$.
\textbf{Inference.}
For the diffusion model, at each time step $t$, we predict the $\hat{x}_{0}$ with the denoising process based on the corresponding multi-modal conditions, then add the noise to $\hat{x}_{0}$ at time step $t-1$ with the diffuse process. 
}
\label{fig:pl}
\end{figure*}

\subsection{Preliminary DDPM} 
Our framework generates motion with a Denoising Diffusion Probabilistic Model (DDPM)~\cite{ho2020denoising} from pure noise sampled from a Gaussian distribution. We follow the DDPM definition of diffusion as a Markov noising process.
We denote the noise motion as $x_t$, where $t$ is a time step and $x_0$ is drawn from the data distribution. 
The forward noising process is defined as :
\begin{equation}
    q\left(x_{t} \mid x_{0}\right)=\mathcal{N}\left(x_t;\sqrt{\bar{\alpha}_{t}} x_{0},\left(1-\bar{\alpha}_{t}\right) I\right) ,
  \label{eq:diff_1}
\end{equation}
where $\bar{\alpha}_{t} \in(0,1)$ are constant hyper-parameters which follow a monotonically decreasing schedule. When $\bar{\alpha}_{t}$ approaches 0, we can approximate $x_{T} \sim \mathcal{N}(0, \boldsymbol{I})$, where $T$ is the total time step.
Our goal is to generate a human motion $\hat{x}_{0}$ given multi-modal conditions $c$. 
We perform the denoising process of gradually cleaning $x_{T}$ by learning a denoising network $D$. 
We follow~\cite{ramesh2022hierarchical, tevet2023human, yang2023diffusestylegesture} to predict the signal itself instead of predicting $\epsilon_{\theta}\left(x_{t}, t\right)$~\cite{ho2020denoising}. 
The network $D$ learns parameters $\theta$ based on the input noise $x_{t}$, noising step $t$ and conditions $c$ to 
reconstruct the original signal $x_{0}$ as :
\begin{equation}
    \hat{x}_{0}= D \left(x_{t}, t, c\right).
\end{equation}
We optimize $\theta$ with the Huber loss for training stability~\cite{huber1992robust} following~\cite{yang2023diffusestylegesture} as :
\begin{equation}
    \mathcal{L}_\mathrm{huber}=E_{x_{0} \sim q\left(x_{0} \mid c\right), t \sim[1, T]}\left[\operatorname{HuberLoss}\left(x_{0}-\hat{x}_{0}\right)\right].
    \label{eq:huberloss}
\end{equation} 

\subsection{Progressive Fusion} 
As shown in Figure \ref{fig:pl}, we consider several different modalities of information, including transcript, speech, motion, identity, and emotion.
Simple multi-modal fusion methods like~\cite{liu2022beat, tevet2023human, yang2023diffusestylegesture} cannot efficiently utilize cross-modal information to model the correspondence between speech and motion.
To this end, we propose a Progressive Fusion Strategy (PFS) including Feature Processing, Specific Feature Encoding, and Shared Feature Encoding.

\noindent\textbf{Feature Processing.} 
Following~\cite{liu2022beat, yang2023diffusestylegesture}, we leverage the pre-trained language model FastText~\cite{bojanowski2017enriching} and the acoustic model WavLM~\cite{chen2022wavlm} model to process transcript and audio, and obtain the transcript feature $\mathbf{e}^{1: N}$ and audio feature $ \mathbf{a}^{1: N}$, respectively, where $N$ is the total frames.
For motion data $x=x^{1: N}$, we employ rotation matrices to record the rotation state of each joint as $x^{i} \in \mathbb{R}^{J \times 9}$, where $i$ represents the $i$-th frame and $J$ is the number of joints.
For convenience, we concatenate transcript and audio features as the speech feature $\mathbf{s}$:
\begin{equation}
    \mathbf{s}=\left[\mathbf{a} || \mathbf{e}\right].
\end{equation}

\noindent\textbf{Masked Style Matrix.} 
Identity and emotion are both significant factors influencing motion styles. 
To effectively control and edit motion generation using both identity and emotion, we unify identity and emotional features with a style matrix $\mathbf{m}_{s}$ inspired by vanilla ``early fusion'' method~\cite{zadeh2017tensor}.
Specifically, $\mathbf{m}_{s}$ can be mathematically equivalent to a differentiable outer product
identity representation $\mathbf{z}_{id}$ and emotion representation $\mathbf{z}_{emo}$ :
\begin{equation}
    \mathbf{m}_{s}=
    \mathbf{z}_{id}
    \otimes
    \mathbf{z}_{emo}.
\end{equation}
Moreover, classifier-free guidance (Section \ref{sec:3_3}) is used to increase the diversity of generated motion and enhance style control.
To this end, a masked style matrix is designed and we leverage its reshaped style token $\mathbf{z}_{s}$ to guide the process of the later feature encoding.

\noindent\textbf{Specific Feature Encoding.} 
One straightforward approach is to directly concatenate speech and motion features as~\cite{liu2022beat, yang2023diffusestylegesture}. 
However, for this ``early fusion'', we assume that it is unnecessary because the $\mathbf{s}$ and $\mathbf{x}$ encompass fine-grained information and high-frequency noise~\cite{nagrani2021attention}.
Specifically, we separately encode the speech feature $\mathbf{s}$ and motion feature $\mathbf{x}$ using individual transformer encoding layers to obtain their specific representations. 
Furthermore, we insert the style token $\mathbf{z}_{s}$ and time token $\mathbf{z}_{t}$ into $\mathbf{s}$ and $\mathbf{x}$. 
This operation can be expressed as:
\begin{equation}
\mathbf{s}'=g\left(\left[\mathbf{z}_{s}|| \sigma\left(\mathbf{s}\right)||\mathbf{z}_{t}\right] +\mathbf{p}\right),
\end{equation}
where $\sigma$ represents linear projection, $g$ represents transformer encoder architecture~\cite{vaswani2017attention} and we use the relative position encoding $\mathbf{p}$~\cite{Kitaev2020Reformer} to maintain temporal effect stability for motion transformations.
Here, the specific speech feature is actually $\mathbf{s}'=\left[\mathbf{z}_{s}^{1}|| \mathbf{s}' ||\mathbf{z}_{t}^{1}\right]$, and we simplify its expression.
We apply the same operation to the motion feature $\mathbf{x}$ to obtain specific motion representations $\mathbf{x}'=\left[\mathbf{z}_{s}^{2}|| \mathbf{x}' ||\mathbf{z}_{t}^{2}\right]$. 
During this process, the style and time tokens also undergo updates. 
The style tokens $\mathbf{z}_{s}^{1}, \mathbf{z}_{s}^{2}$ learn the style representations of specific features, while the time tokens $\mathbf{z}_{t}^{1}, \mathbf{z}_{t}^{2}$ constrain the specific feature dependencies on noisy time steps.
Different modal features learn their own characteristics without being influenced by other modalities by utilizing progressive fusion.
Moreover, the style matrix $\mathbf{m}_{s}$ effectively learns specific features from different modalities, facilitating control and editing of motion generation.

\noindent\textbf{Shared Feature Encoding.} 
Under the guidance of specific style tokens and time tokens, we use cross-attention~\cite{vaswani2017attention} to merge the specific speech feature $\mathbf{s}'$ and motion feature $\mathbf{x}'$ as:
\begin{equation}
\label{eq:r3_1}
\mathbf{f}=\operatorname{SoftMax}\left( \mathbf{x}' \left( \mathbf{s}' \right)^{T} / \sqrt{d}\right) \mathbf{s}' \text {,}
\end{equation}
where $\mathbf{f}$ and $d$ represents the shared feature and the feature dimensions, respectively. $\mathbf{z}_{s}^{3}$ and $\mathbf{z}_{t}^{3}$ are shared style and time token, respectively. 
To fully leverage meaningful variance among different modality features, we aggregate both specific and shared features to jointly influence motion generation, enhancing accuracy and diversity. 
Additionally, we add different style tokens and time tokens to obtain fused token $\mathbf{z}_{f}$, which is stacked into the fused feature sequence to maintain training stability and consistency.
Similarly, we use the concatenated hybrid feature  $\mathbf{f}'$ as input to the transformer encoding layers. 
This process can be represented as :
\begin{equation}
\mathbf{f}'=g\left(\left[\left(\mathbf{z}_{s}^{i} + \mathbf{z}_{t}^{i}\right) || \sigma\left(\mathbf{x}'|| \mathbf{s}' || \mathbf{f}\right)\right]\right),
\end{equation}
where $i=1,2,3$, represents specific and shared tokens in two feature encoding stages.
Finally, $\mathbf{f}'$ is mapped to the same dimension as $\mathbf{x}_{0}$ after a linear layer.
The fusion of specific and shared features allows cross-modal information to be exchanged between features while strengthening the connection between the final output and style matrix $\mathbf{m}_{s}$, facilitating motion editing.

\noindent\textbf{Geometric Loss.}
Geometric loss can be employed in kinematic motion generation to enhance the physical realism, generating natural and coherent motion~\cite{shi2020motionet, tevet2023human, tseng2023edge}. 
In addition to signal-based supervision $\mathcal{L}_{huber}$ in Equation \ref{eq:huberloss},  we utilize joint velocity and acceleration loss to supervise motion generation:
\begin{equation}
    \mathcal{L}_{\mathrm{vel}}=\frac{1}{N-1} \sum_{j=1}^{N-1}\left\|\left(x_{0}^{j+1}-x_{0}^{j}\right)-\left(\hat{x}_{0}^{j+1}-\hat{x}_{0}^{j}\right)\right\|_{2}^{2} ,
\end{equation}

\begin{equation}
    \mathcal{L}_{\mathrm{acc}}=\frac{1}{N-2} \sum_{j=1}^{N-2}\left\|\left(v_{0}^{j+1}-v_{0}^{j}\right)-\left(\hat{v}_{0}^{j+1}-\hat{v}_{0}^{j}\right)\right\|_{2}^{2},
\end{equation}
where $j$ represents the $j$-th frame, and $v_{0}^{j}=x_{0}^{j+1}-x_{0}^{j}$. 
We expand the supervision of joint acceleration based on the velocity loss in~\cite{tevet2023human}, which contributes to generating smoother and more natural motion. 
The overall training loss can be represented as follows:
\begin{equation}
\mathcal{L}=\mathcal{L}_{\mathrm{huber}}+\lambda_{\mathrm{vel}} \mathcal{L}_{\mathrm{vel}}+\lambda_{\mathrm{acc}} \mathcal{L}_{\mathrm{acc}},
\end{equation}
where $\lambda_{\mathrm{vel}}$ and $\lambda_{\mathrm{acc}}$ are set to 0.1 and 0.01, respectively in our experiments.

\begin{figure}[!t]
    \centering
    \begin{minipage}{0.8\linewidth}
      \centerline{\includegraphics[width=9cm]{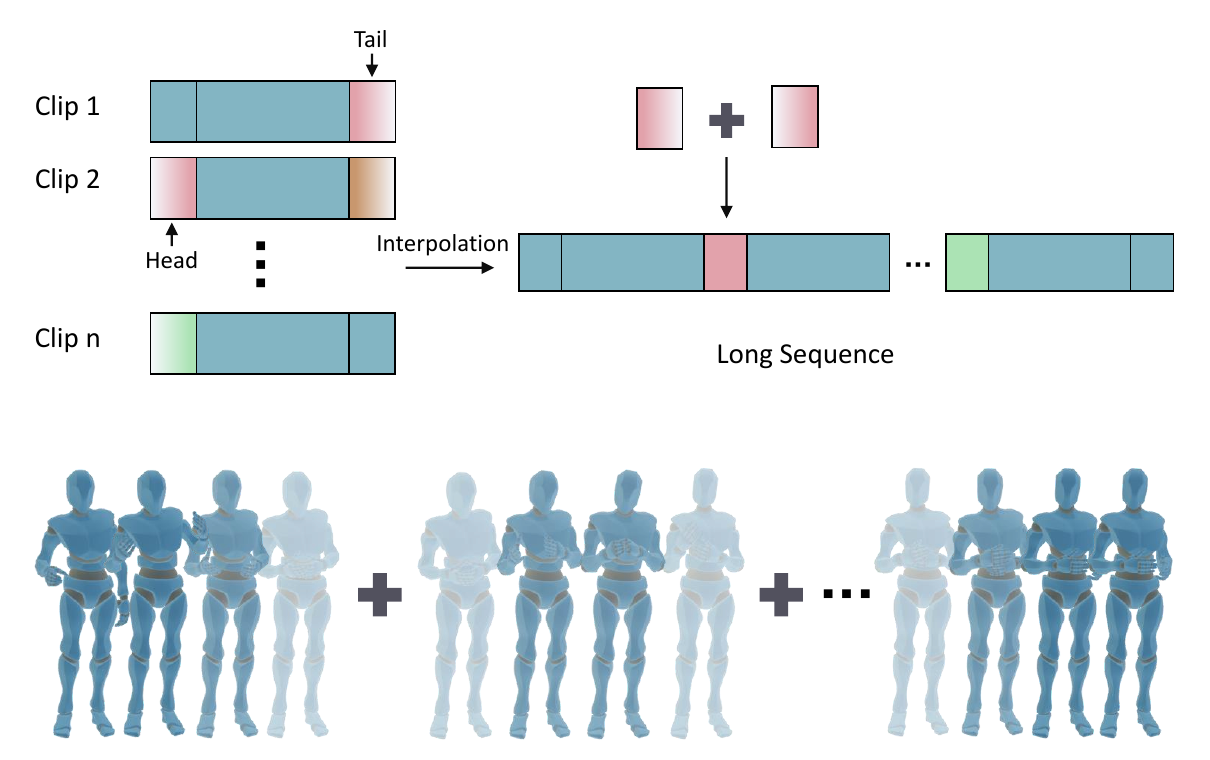}}
    \end{minipage}
\caption{We train MMoFusion on 10-second clips and generate motion of any length by interpolating the tail of the previous motion clip and the head of the next motion clip, as represented by the overlapped clips in the batch that share the same color.}
\vspace{-1.5em}
\label{fig:ls}
\end{figure}

\subsection{Sampling} 
\label{sec:3_3}
In every time step $t$, we predict the denoised sample $\hat{x}_0$ and add the noise to time step $t-1$ according to Equation \ref{eq:diff_1}, terminating when $t = 0$. 

\noindent\textbf{Classifier-free Guidance.}
Following~\cite{tevet2023human, yang2023diffusestylegesture}, we implement classifier-free guidance~\cite{ho2021classifier} for our denoising network $D \left(x_{t}, t, c\right)$ to trade-off diversity and fidelity of the generated motion. 
Specifically, our multi-modal conditions $c$ including style matrix $\mathbf{m}_s$ formed by identity $\mathbf{z}_i$ and emotion $\mathbf{z}_e$. And the speech features $\mathbf{s}$ formed by audio $\mathbf{a}$ and transcript $\mathbf{e}$. 
We randomly replace the style matrix $\mathbf{m}_s$ with $\emptyset$ in low probability (\eg{,} $10\%$). 
However, we argue that applying the random mask to the speech feature $\mathbf{s}$ is not an ideal balance between diversity and fidelity due to the temporal information in $\mathbf{s}$ as shown in Table \ref{tab:ab_3}.
To this end, we achieve classifier-free guided motion generation by combining the outputs of the conditional diffusion model $D \left(x_{t}, t, \left(\mathbf{m}_s, \mathbf{s}\right)\right)$ and unconditional diffusion model $D \left(x_{t}, t, \left(\emptyset, \mathbf{s}\right)\right)$ during sampling.
\begin{equation}
    \hat{x}_{0}=\omega  D \left(x_{t}, t, \left(\mathbf{m}_s, \mathbf{s}\right)\right)+(1-\omega ) D\left(x_{t}, t, \left(\emptyset, \mathbf{s}\right)\right),
\end{equation}
where $\omega$ is a guidance weight to scale conditions.

\noindent\textbf{Long Sequences Sampling.}
In most cases, co-speech-generated motion sequences can be of arbitrary length. 
However, the diffusion model generates fixed-length motion sequences.
Some methods generate multiple motion segments separately and concatenate them~\cite{zhu2023taming, yang2023diffusestylegesture} to synthesize coherent motion sequences aligned with the duration of the speech. 
However, this is time-consuming and unstable for generating continuous motion.
To address this, as shown in Figure \ref{fig:ls}, we split the speech signal into a few overlapped clips and imposed temporal constraints on batches of sequences. 
Additionally, to eliminate discontinuities between generated motion clips, we apply linear interpolation between the tail of the previous motion clip and the head of the next motion clip within a batch. 
This ensures the continuous generation of long motion sequences. 
In practice, we select 30 frames of head or tail clip to apply linear interpolation. 
Our sampling method takes 42 minutes to evaluate the test set, while the previous methods~\cite{zhu2023taming, yang2023diffusestylegesture} take 407 minutes. 

\section{Experiments}
\subsection{Datasets and Experimental Setting}
\noindent\textbf{Datasets.}

\textit{1) BEAT}~\cite{liu2022beat} is a large-scale multi-modal human motion dataset. 
It contains 60 hours of English-speaking data, including 30 speakers with 8 different emotions.  
For each motion, it provides audio, transcripts, identity, and emotion labels. 
In the dataset, 10 speakers contribute 4 hours each, while the other 20 provide 1 hour each. 
To establish a benchmark for training consistency, we select data from 30 speakers, contributing 1h each. 
This ensures that the transcripts are consistent for each speaker.
We split data for training, validation, and testing by approximately 70\%, 10\%, and 20\%. 

\textit{2) TED Expressive} ~\cite{liu2022learning}: The TED Expressive dataset includes both finger and body motions. A 3D pose estimator is used to capture pose information in the data. TED Expressive annotates the 3D coordinates of 43 key points, including 13 upper body joints and 30 finger joints.

\noindent\textbf{Evaluation Metrics.}
We use the common four metrics to evaluate the performance: 
(i) \textit{Fréchet Gesture Distance (FGD)} 
measures the distance between the generated motion distribution and the real motion data distribution\cite{yoon2020speech}. 
Following~\cite{liu2022beat}, we train an auto-encoder as a pre-trained model to extract the synthetic motion features and real motion features.
(ii) \textit{Diversity} evaluates the variations among generated motion corresponding to various inputs~\cite{lee2019dancing}. 
We utilize the FGD auto-encoder to obtain latent features from motion and calculate the mean feature distance.
Following~\cite{zhu2023taming} we randomly select 500 synthetic motions and input them to FGD auto-encoder to compute the average absolute error between features.
(iii) \textit{Semantic-Relevant Gesture Recall (SRGR).} 
BEAT\cite{liu2022beat} employs semantic relevance scoring as weights for the Probability of Correct Keypoint (PCK) between generated motion and ground truth motion. 
Where PCK is the number of joints successfully recalled against a specified threshold $\delta$.
(iv) \textit{Beat Alignment Score (BeatAlign)}
\cite{li2021ai} evaluates the rhythmic synchrony between generated motion and audio by computing the distance between motion kinematic beats and audio beats.

\noindent\textbf{Implementation Details.}
We downsample the motion data to 30 frames per second (fps) and segment them into several clips with max-length 300 frames for training.
We also downsample the audio to 16kHz and use linear interpolation to align the extracted WavLM\cite{chen2022wavlm} feature with motion sequence in the time dimension. 
Besides, to further utilize the audio information, we also use the MFCC, mel-scaled spectrogram, prosody, and onset feature.
These features are concatenated as the audio feature representation $\mathbf{a} \in \mathbb{R}^{1133}$.
The transcript feature $\mathbf{e} \in \mathbb{R}^{301}$ is obtained by a pre-trained language model FastText~\cite{bojanowski2017enriching}.
In our experiments, the dimensions of speech and motion latent features are 96, and 384 respectively.
The diffusion step is 1000, and we train the overall framework for 120K iterations with a batch size of 150 on one NVIDIA 4090 GPU.

\begin{table}[!t]
  \centering
  \caption{Comparison with the state-of-the-art methods, $\omega$ refers to the guidance weight at sampling. \textbf{Bold} represent optimal result. $\dagger$ indicates the results from their papers.}
    \renewcommand{\arraystretch}{1.2}
    \resizebox{\linewidth}{!}{
        \begin{tabular}{cc|cccc|cccc}
        \hline
       \multirow{2}{*}{}  & \multirow{2}{*}{\centering Method} & \multicolumn{4}{c|}{Modalities}  & \multirow{2}{*}{FGD $\downarrow$} & \multirow{2}{*}{Diversity $\uparrow$} & \multirow{2}{*}{SRGR $\uparrow$} & \multirow{2}{*}{BeatAlign $\uparrow$} \\
        \cline{3-6}     &    & Audio & Text & Emo &{ID } &       &       &       &  \\
        \hline
        \multirow{5}[4]{*}{\rotatebox{90}{Upper Body}} & CaMN\cite{liu2022beat} &\checkmark &\checkmark &\checkmark & \checkmark &  18.9  & 53.2  & \textbf{0.217}  & 0.843 \\
        & MDM\cite{tevet2023human} &\checkmark &\checkmark & &  &  34.1  & 55.7  & 0.209  & 0.788 \\
        & DSG\cite{yang2023diffusestylegesture} &\checkmark &\checkmark & & \checkmark & 48.4  & 61.4 & 0.214 & 0.841 \\ 
        & MambaG$^\dagger$\cite{fu2024mambagesture} &\checkmark &\checkmark &\checkmark & \checkmark & -  & - & 0.213 & \textbf{0.863} \\ 
        \cline{2-10}
        & Our ($\omega=1$)  & \checkmark & \checkmark& \checkmark&\checkmark & \textbf{12.0}  & 66.5 & 0.215 & 0.836 \\
        & Our ($\omega=3$) & \checkmark& \checkmark& \checkmark&\checkmark & 12.5  & \textbf{72.1}  & 0.215 & \textbf{0.845} \\
        \hline
        \multirow{5}[4]{*}{\rotatebox{90}{Full Body}} & CaMN\cite{liu2022beat}  & \checkmark & \checkmark& \checkmark&\checkmark &  3.5  & 83.7  & 0.239  & 0.834 \\
        & MDM\cite{tevet2023human}  &\checkmark &\checkmark & & &  5.1  & 83.4  & 0.234  & 0.748 \\
        & DSG\cite{yang2023diffusestylegesture} &\checkmark &\checkmark & & \checkmark& 31.3  & 101.2 & 0.238 & 0.837 \\
        & MambaG$^\dagger$\cite{fu2024mambagesture} &\checkmark &\checkmark &\checkmark & \checkmark & -  & - & 0.237 & \textbf{0.853} \\ 
        \cline{2-10}
        & Our ($\omega=1$)  & \checkmark & \checkmark& \checkmark&\checkmark  & \textbf{0.5}  & 95.0  & \textbf{0.240} & 0.827 \\
        & Our ($\omega=3$) & \checkmark & \checkmark& \checkmark&\checkmark  & 1.2  & \textbf{104.4}  & \textbf{0.240} & 0.839 \\
        \hline
        \end{tabular}%
    }
  \label{tab:qc_1}%
\end{table}%

\subsection{Quantitative Results}
We compare our method with three state-of-the-art co-speech motion generation approaches. 
To ensure a fair comparison in the experiments, we retrained CaMN\cite{liu2022beat}, MDM\cite{tevet2023human}, and DSG\cite{yang2023diffusestylegesture} on our benchmark and report the result of MambaGesture\cite{fu2024mambagesture} from their paper.
We also retrain DiffGesture~\cite{zhu2023taming} but did not converge in our experiments with the same settings as~\cite{yin2023emog} mentioned. 
This may be attributed to the motion data format is different. 
Consequently, we train our model on the TED Expressive to compare with other methods including DiffGesture.

The quantitative comparison results are shown in Table \ref{tab:qc_1}. 
Our method outperforms existing methods on both upper body and challenging full-body motion generation.
Our approach leverages multi-modal information and classifier-free guidance to generate motion, ensuring high-fidelity and diverse motion. 
We focus on the fusion of multi-modal information, not just its accuracy, as it can restrict the generation of results. 
Additionally, the randomness in diffusion model sampling can also adversely affect the SRGR metric since it focuses solely on the accuracy between results and ground truth. 
However, generating vivid motion may not necessarily require consistency with the ground truth.

We also test the model performance on the TED Expressive. Our method still achieves superior performance in Table \ref{tab:qc_2}, indicating its enhanced generalization compared to DiffGesture, which cannot converge on the BEAT dataset~\cite{yin2023emog}.

\begin{table}[h]
  \centering
  \caption{Compare with SOTA method on TED Expressive.}
    \renewcommand{\arraystretch}{1.2}
    \small
        \begin{tabular}{cccc}
        \hline
        Method & FGD $\downarrow$  & BeatAlign $\uparrow$ & Diversity $\uparrow$ \\
        \hline
        HA2G \cite{liu2022learning}   & 5.3   & 0.641  & 173.9 \\
        LivelySpeaker (RAG)\cite{zhi2023livelyspeaker}  & 4.5   & 0.714  & 181.6 \\
        DiffGesture\cite{zhu2023taming}  & \textbf{2.6}   & 0.718  & 182.8 \\
        Ours &2.8   &\textbf{0.725}  &\textbf{184.3}  \\   \hline
        \end{tabular}%
  \label{tab:qc_2}%
\end{table}%

\begin{figure}[!t]
    \centering
    \includegraphics[width=1\linewidth]{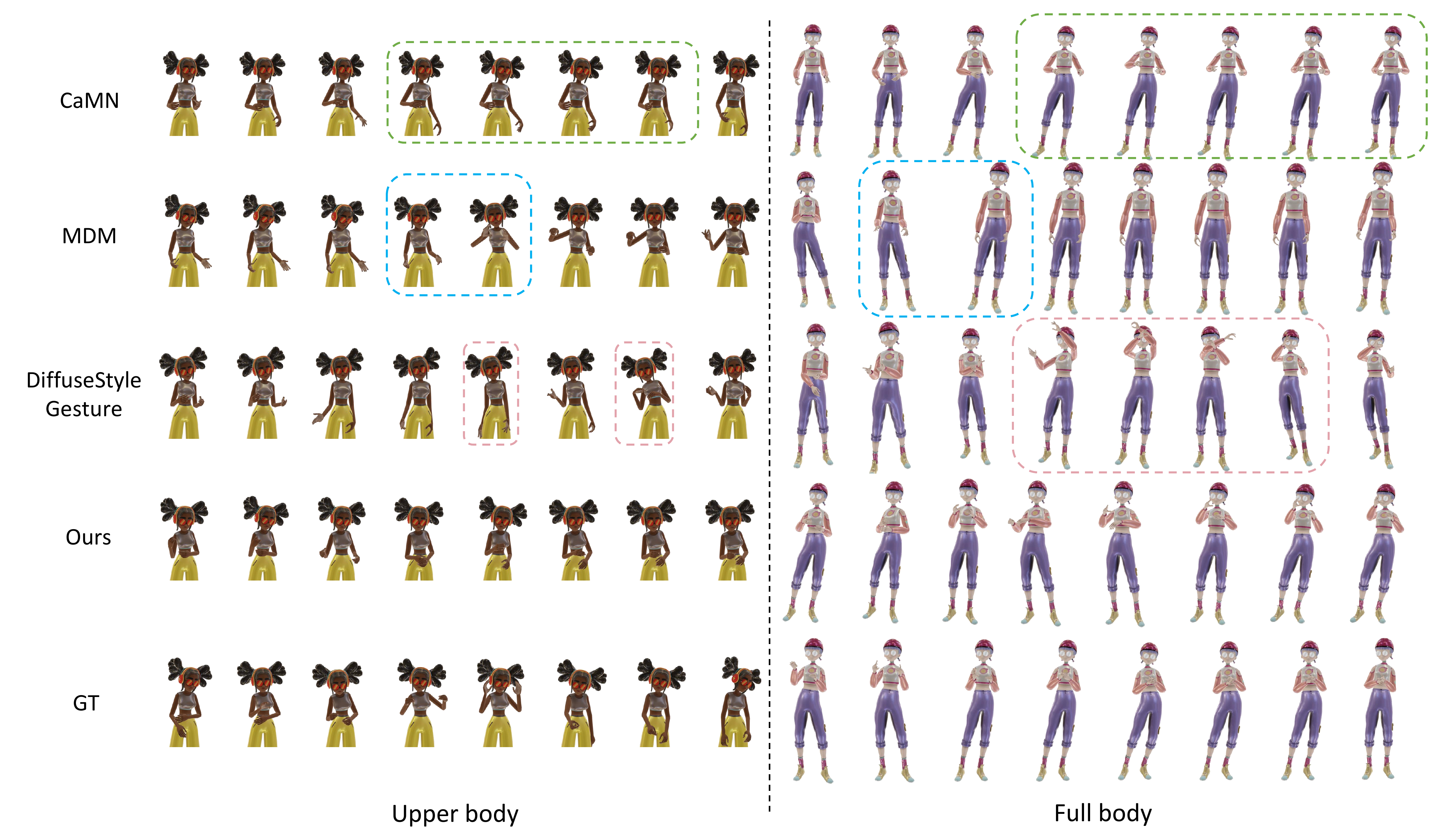}
    \caption{
    Visual comparisons of upper body and full body motion generation results.
    }
    \label{fig:ex_1}
\end{figure}

\subsection{Qualitative Results}
\noindent\textbf{Comparison with state-of-the-art methods.} 
We compared both the upper body and full body results of our method with other state-of-the-art methods.
From Figure \ref{fig:ex_1}, we can see that 
CaMN\cite{liu2022beat} use a simple multi-modal feature concatenation that leads to the generation of monotonous motion (green boxes).
MDM\cite{tevet2023human} generates motion by token information which lacks temporal expression resulting in discontinuous (blue boxes).
DSG\cite{yang2023diffusestylegesture} relies on attention, leading to unreasonable results (orange boxes). 
In contrast, our method can generate diverse and vivid motion.

\begin{figure}[!t]
    \centering
    \includegraphics[width=0.9\textwidth]{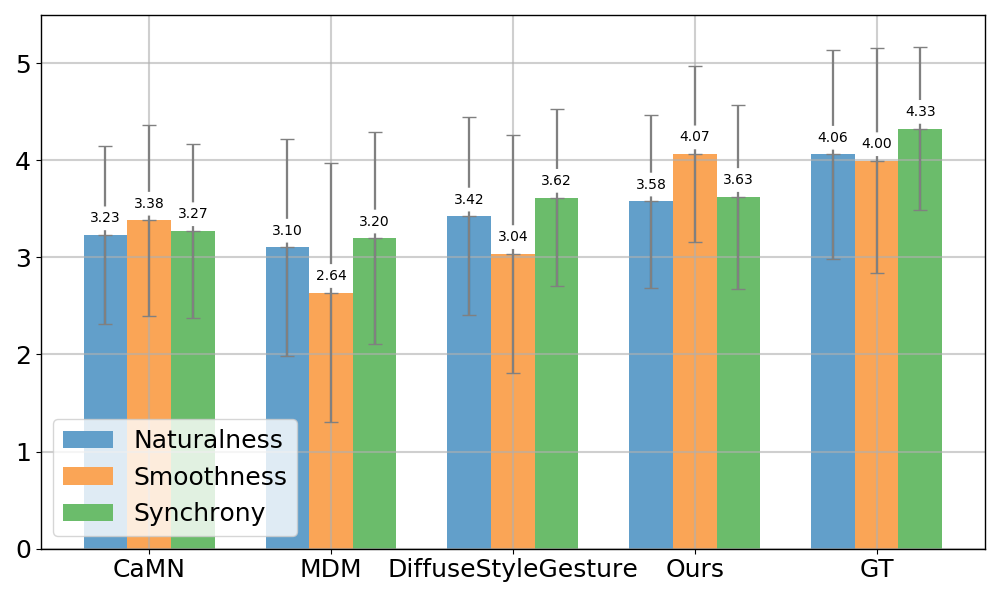}
    \caption{User study on naturalness, smoothness, and synchrony. 
    Error bars represent the standard deviation.}
    \label{fig:us_1}
\end{figure}

\noindent\textbf{User study.}
Since quantitative comparisons may not accurately assess the quality of generated motion~\cite{yoon2020speech}, we conducted a user study with 18 recruited volunteers. 
Following~\cite{liu2022learning, zhu2023taming}, we objectively evaluate the naturalness, smoothness, and synchrony of the generated motion. 
The scoring ranges from 1 to 5, with higher scores indicating better quality.
As shown in Figure \ref{fig:us_1}, our method outperforms the state-of-the-art methods in all three metrics. 
Particularly, due to errors in data collection, our approach surpasses even real data in terms of smoothness, which indicates the effectiveness of our proposed framework.

\begin{figure}[!t]
    \centering
    \begin{minipage}{0.9\linewidth}
        \centering
         \subfloat[\rmfamily{Emotion control.}]{
                \centering
              \includegraphics[width=0.48\textwidth]{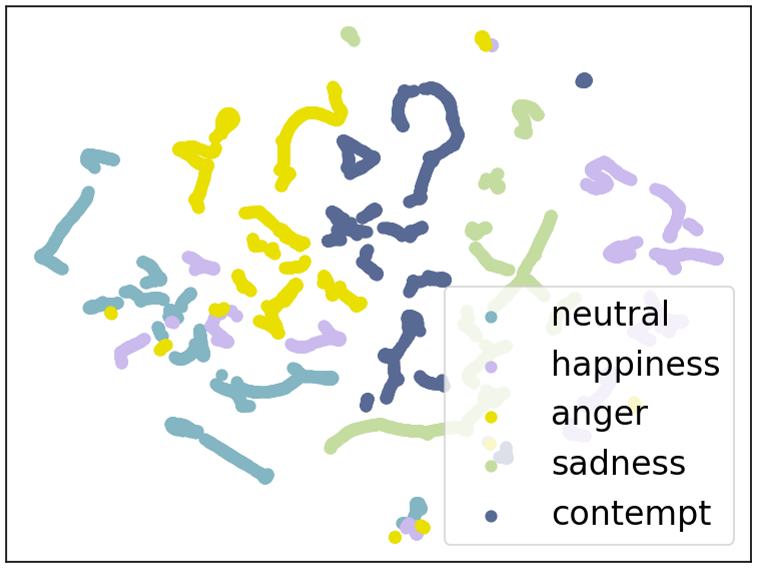 }
        }
         \subfloat[\rmfamily{Identity control.}]{
                \centering
              \includegraphics[width=0.48\textwidth]{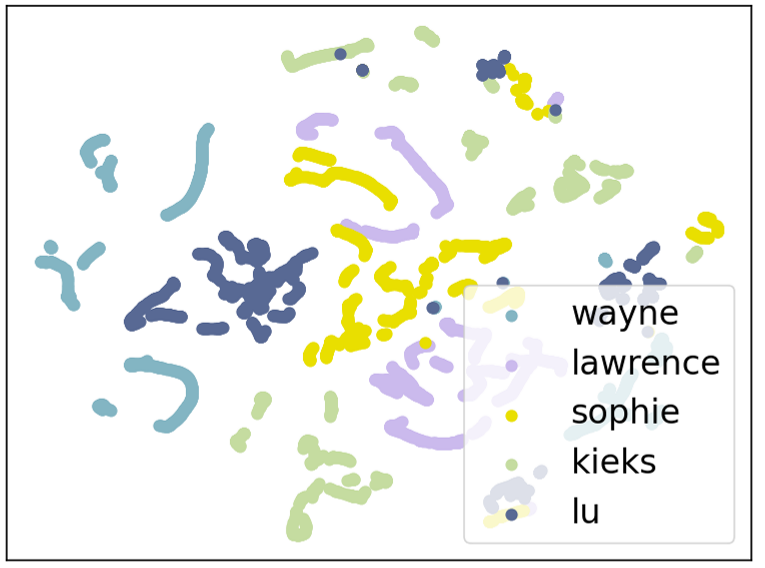}
        }
        \caption{T-SNE visualization results in five different emotions and identities by inputting the same speech.
        Motion with different styles is basically distinguished into different groups.}
        \label{fig:sc_1}
    \end{minipage}
\end{figure}

\begin{figure}[!t]
    \centering
      \includegraphics[width=0.96\linewidth]{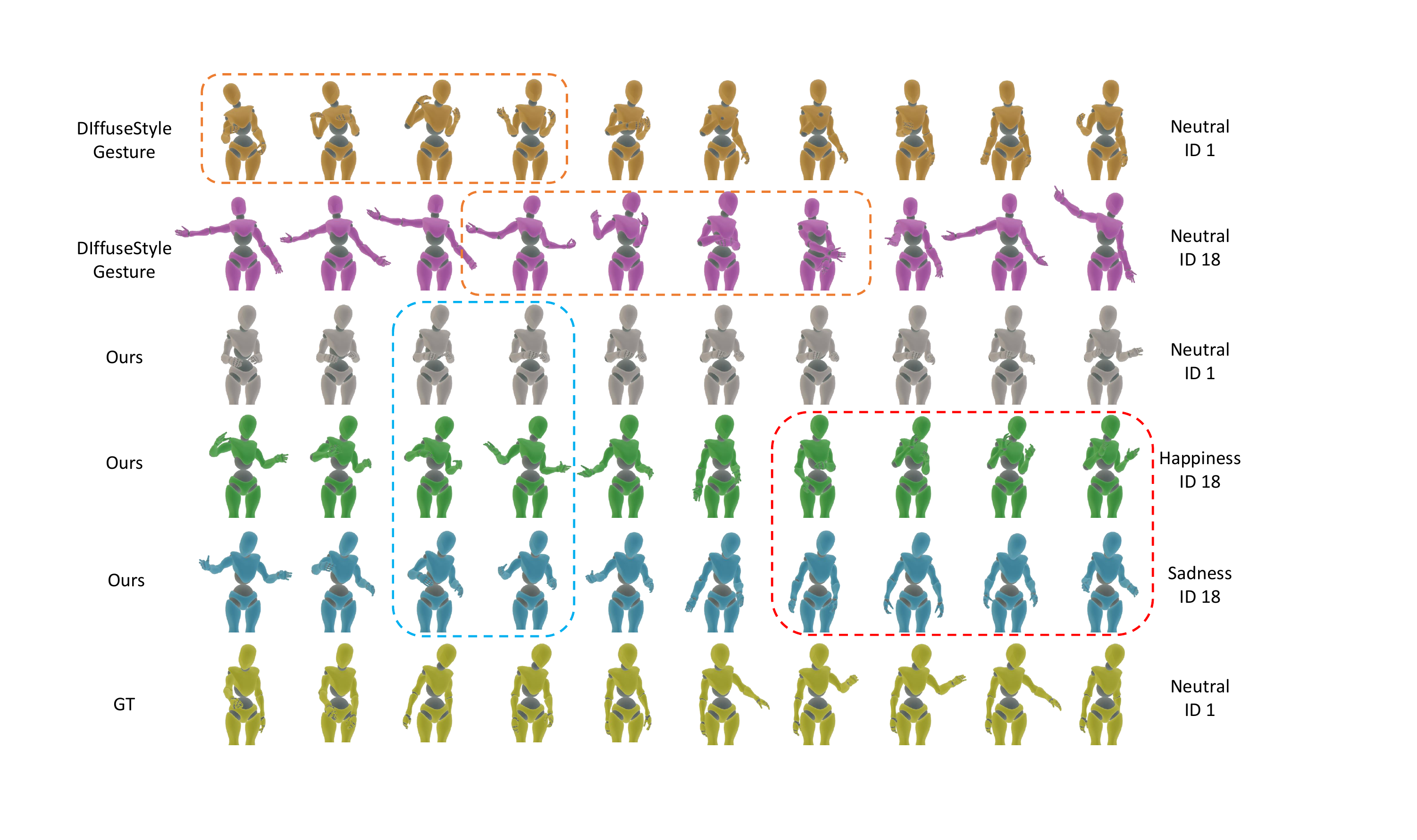}
    \caption{Visualization results of style control. 
    Motion generated by DSG is \textcolor[RGB]{237,125,49}{unreasonable}, and our framework generates more \textcolor[RGB]{91,155,213}{vivid} and \textcolor[RGB]{91,155,213}{diverse} motion compared to GT by controlling styles, which also exhibits clear \textcolor[RGB]{255,0,0}{semantic} differences.}
\label{fig:sc_2}
\end{figure}

\noindent\textbf{Style control.}
We utilize a masked style matrix during progressive fusion and further guide motion generation, which enables free control over the style of generated motion. 
To illustrate this, we input a speech and use different emotions and identity information separately to generate motion sequences. 
The T-SNE visualization results are shown in Figure \ref{fig:sc_1}.
To demonstrate the effect of style control, we use the same audio and generate several different motion sequences by editing different identities and emotional inputs. 
We also compare our results with DSG~\cite{yang2023diffusestylegesture}, which also supports identity editing. 
As shown in Figure \ref{fig:sc_2}, DSG generates unreasonable motion as we mentioned in qualitative results. 
While our framework generates more diverse and vivid motion than the ground truth. 
Additionally, the control operations exhibit clear semantic differences; for example, motion during happy states is inherently more positive compared to those during sad states.

\noindent\textbf{Custom speech.}
Thanks to pre-trained acoustic and language models, our framework supports custom speech input and editing of identities and emotions to generate vivid human motion.
We showcase custom-generated motion using speech ‘in the wild’ in the supplementary material.

\subsection{Model Analysis}
\textbf{Ablation Study on the proposed components.} 
To evaluate the effectiveness of the proposed PFS, we simply concatenate multi-modal features including motion, speech, transcript, emotion, and identity as input to transformer encoding layers to establish a baseline, named ``early fusion''. 
As shown in Table \ref{tab:ab_1}, firstly, we utilize a masked style matrix to guide the motion generation and improve the quality of motion ($2^{th}$ row). 
To reduce the effect of redundant information,
we design a specific feature encoding to encode speech and motion features ($3^{th}$ row), separately.
Moreover, the shared feature encoding involves a cross-attention to aggregate the specific features and obtain the hybrid feature ($4^{th}$ row).
The results show that compared with early fusion, the proposed progressive fusion strategy helps the model extract key information to model the mapping relationship between speech and motion features.
To generate more physically realistic motion, we employ a geometric loss including joint velocity and acceleration ($5^{th}$ row). 
Finally, we use classifier-free guidance (CFG) to generate more diverse motion ($6^{th}$ row).
We can also see the geometric loss impacts motion speed, slightly reducing rhythm-matching accuracy (BeatAlign) but enhancing motion fidelity (FGD) ($7^{th}$ row).

\begin{table}[!t]
  \centering
  \caption{Ablation Study on the proposed components. $\mathcal{L}_{geo}$ stands for the geometric loss and CFG stands for classifier-free guidance.
    \textbf{Bold} represents optimal results.}
    \renewcommand{\arraystretch}{1.2}
        \begin{tabular}{cccccc}
        \hline
         No.& Setting & FGD $\downarrow$  & Diversity $\uparrow$ & SRGR $\uparrow$  & BeatAlign $\uparrow$ \\
        \hline
        1& Early Fusion & 19.3   & 66.2 & 0.214  & 0.832 \\
        2& + Style Matrix & 18.7  & 62.2 & 0.214  & 0.836 \\
        3& + Specific Feature & 14.0  & 64.4  & \textbf{0.215} & 0.838 \\
        4& + Shared Feature & 13.9  & 66.8  & \textbf{0.215}  & 0.839 \\
        5& + $\mathcal{L}_{geo}$ & \textbf{12.0}  & 66.5 & \textbf{0.215} & 0.836 \\
        6& + CFG ($\omega = 3$) & 12.5  & 72.1 & \textbf{0.215}  & 0.845 \\ 
        7& -- $\mathcal{L}_{geo}$   &  17.7  &69.2  & 0.214  & \textbf{0.847} \\
        \hline
        \end{tabular}%
  \label{tab:ab_1}%
\end{table}%

\noindent \textbf{Ablation study on the progressive fusion strategy.}
Since the granularity of progressive fusion is controllable, we conduct an ablation experiment on the number of transformer encoding layers used for specific feature encoding and shared feature encoding. 
As shown in Table \ref{tab:ab_2}, the best performance is achieved when the number of layers in specific feature encoding is 4 and in shared feature encoding is 2.

\begin{table}[h]
    \centering
    \caption{Ablation Study on the progressive fusion strategy. $n$ and $m$ are the layer numbers in specific feature encoding and shared feature encoding, respectively.}
      \renewcommand{\arraystretch}{1.2}
        \begin{tabular}{c|c|cccc}
            \hline
        $n$   & $m$   & FGD $\downarrow$  & Diversity $\uparrow$ & SRGR $\uparrow$  & BeatAlign $\uparrow$ \\
            \hline
        3     & 3     & 16.9  & 61.0    & 0.2149 & \textbf{0.845} \\
        4     & 2     & \textbf{12.0}    & \textbf{66.5}  & \textbf{0.2151} & 0.836 \\
        5     & 1     & 14.0    & 63.5  & 0.2150 & 0.836 \\
            \hline
        \end{tabular}%
      \label{tab:ab_2}%
\end{table}

\begin{figure}[h]
    \centering
    \includegraphics[width=0.6\linewidth]{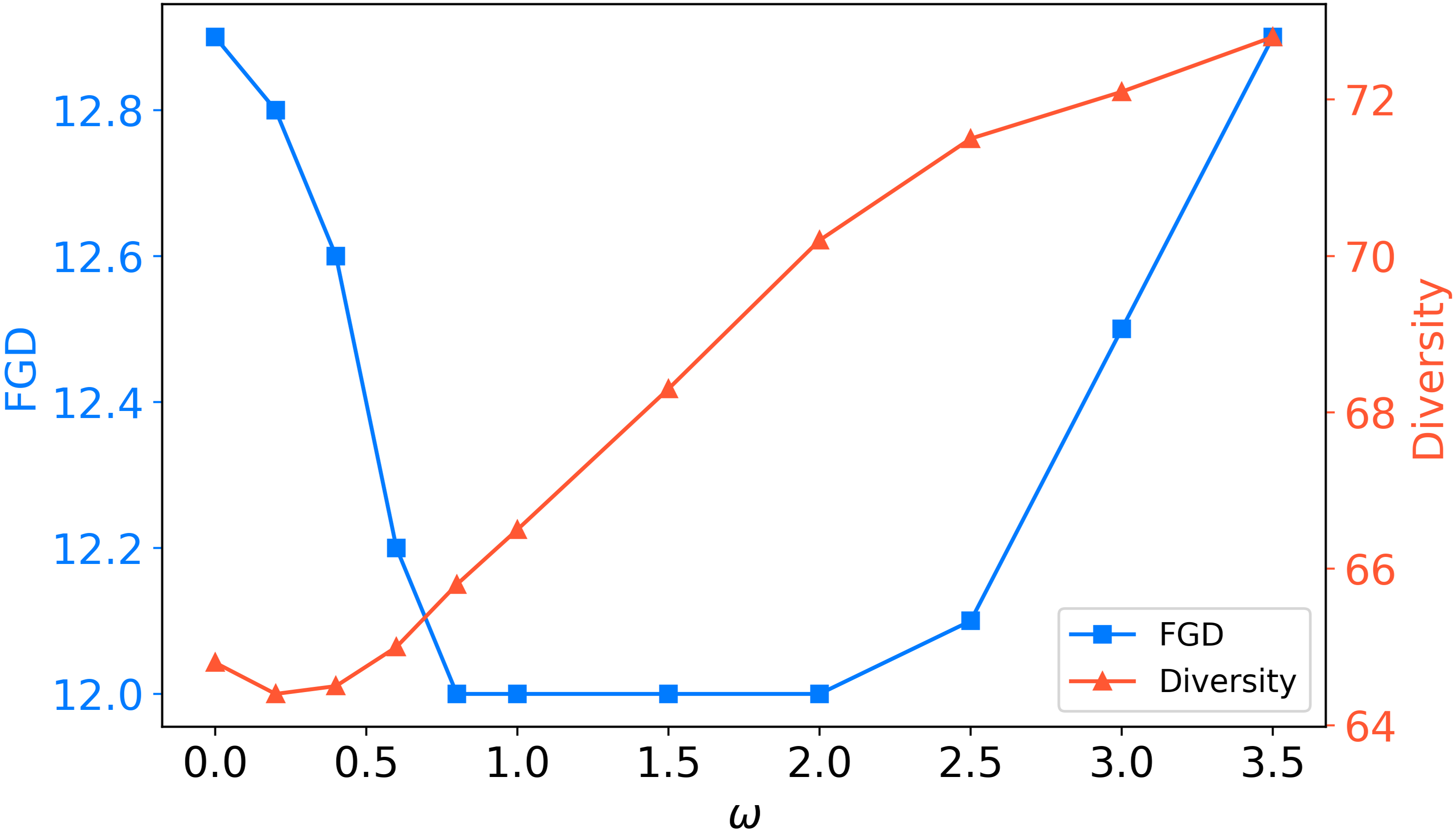}
    \caption{Ablation study on the classifier-free guidance. $\omega$ refers to the guidance weight.}
    \label{fig:ab_4}
\end{figure}

\noindent \textbf{Ablation study on the classifier-free guidance.}
We use classifier-free guidance by employing a random mask on the style matrix to enhance the diversity of motion.
However, as shown in Table \ref{tab:ab_3}, imposing a random mask on the speech feature as one of the conditions destroys the temporal information of the feature and cannot generate high-fidelity motion. We also evaluate the effect of the guidance weight $\omega$ on the generated motion. 
As shown in Figure \ref{fig:ab_4}, too small/large $\omega$ cannot achieve optimal performance.

\begin{table}[h]
    \centering
    \caption{Ablation study on the random mask during classifier-free guidance.}
      \renewcommand{\arraystretch}{1.2}
        \begin{tabular}{c|cccc}
            \hline
            & FGD $\downarrow$  & Diversity $\uparrow$ & SRGR $\uparrow$  & BeatAlign $\uparrow$ \\
            \hline
        Ours    &\textbf{12.0}    & \textbf{66.5}  & \textbf{0.2151} & \textbf{0.836} \\
        w/ speech mask   &22.2    & 60.3  & 0.2132 & 0.827 \\
            \hline
        \end{tabular}%
      \label{tab:ab_3}%
\end{table}

\noindent \textbf{Efficiency of progressive fusion and model complexity discussion.}
To illustrate the effectiveness of proposed progressive fusion, we discuss the parameters, complexity, and time required for the inference test set of the current SOTA method in Table \ref{tab:2}. 
Our method is relatively efficient, but compared with CaMN, the inference speed of the diffusion model needs to be improved.

\begin{table}[h]
  \centering
  \caption{Model complexity analysis.}
  \small
        \begin{tabular}{cccc}
        \hline
        Method & Params (M)  & FLOPs (G) & Time (Min) \\
        \hline
        CaMN   & 50.3  & 15.1  & \textbf{14} \\
        DSG & 8.9  & 2.5  & 407 \\
        Ours  & \textbf{8.3}  & \textbf{2.3} & 42 \\   \hline
        \end{tabular}%
  \label{tab:2}%
\end{table}%

\section{Conclusion}
In this paper, we propose MMoFusion, a multi-modal co-speech motion generation framework with a diffusion model. 
To integrate vastly different modalities effectively, we propose a Progressive Fusion Strategy. 
Specifically, we utilize a masked style matrix that interacts with identity and emotion information to guide motion generation and further control the motion styles.
To model the many-to-many matching relationships between speech and motion temporally, we propose specific feature encoding and shared feature encoding to extract specific and shared features and further merge them.
We also introduce geometric loss including joint velocity and acceleration to smooth motion sequences. 
Besides, to overcome the fixed sequence constraints imposed by the diffusion model, we design a long sequence sampling to generate motion of arbitrary length.
Extensive experiments demonstrate that our framework can produce coherent, realistic, and diverse upper-body and full-body motion, outperforming existing co-speech motion generation methods.

\section{Limitations and Future Work}
Our framework does not account for the complex character displacement involved in full-body motion. 
Consequently, the generated full-body motion may exhibit positional biases, affecting the overall visual perception.
Moreover, co-speech motion generation may raise ethical concerns, including privacy issues, societal biases, and the risk of technical deception. 
Addressing these concerns is crucial to ensure responsible development and deployment of such technologies.

\appendix

\section{Implementation Details}
\noindent \textbf{Feature Processing.}
For transcript, we follow BEAT\cite{liu2022beat} to use Montreal Forced Aligner \cite{mcauliffe2017montreal} to align transcript and audio, so that it also has timing characteristics.
Additionally, the identity and emotion are embedded to the identity representation $\mathbf{z}_{i}$ and emotion representation $\mathbf{z}_{e}$, respectively by a embedding layer.
The time step $t$ and noisy motion data $x$ are projected to the time token $\mathbf{z}_{t}$ and motion feature $\mathbf{x}$, respectively. 
We train motion data with 61 joints of the upper body and 75 of the full body.

\noindent \textbf{User Study.}
We randomly selected 6 speeches in the test set and develop a rating interface shown in Figure \ref{fig:ex_1} to score the motion videos generated by different methods.
The participants do not know in advance the corresponding generation method of the motion videos, which ensures the reliability of the evaluation results.
The participants are almost all students from our lab and a few from other schools, aged between 20 and 30.

\begin{figure*}[!t]
    \centering
    \includegraphics[width=0.98\linewidth]{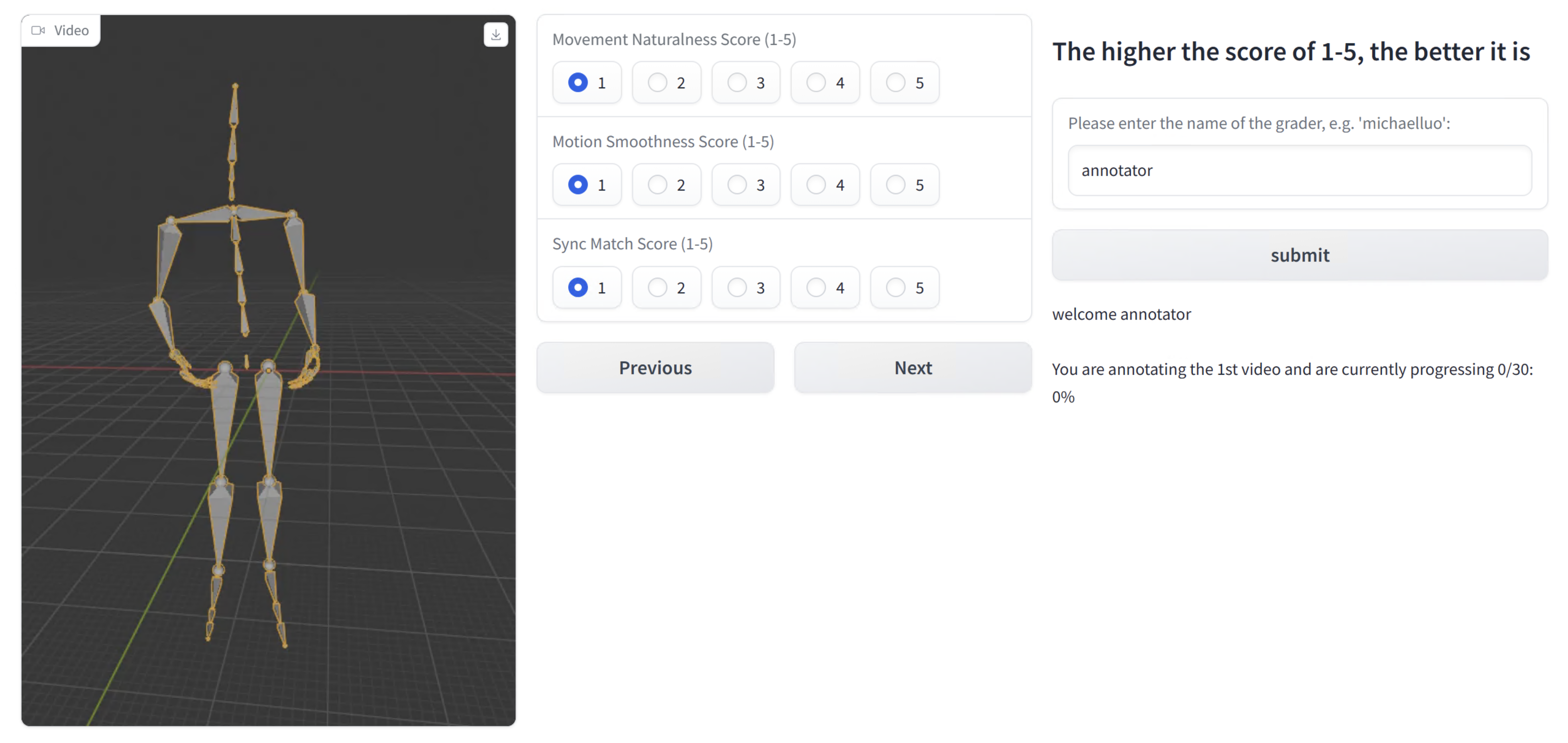}
    \caption{
    Screenshot of the rating interface from the user study.}
    \label{fig:ex_1}
\end{figure*}

\section{Supplementary Video}
\noindent \textbf{Visual Comparisons.}
We utilize free characters in Mixamo \footnote{https://www.mixamo.com/} to visualize the generated motion sequences by Blender \footnote{https://www.blender.org/}.
Please refer to \textit{supplementary video} for more intuitive comparison.
The visual comparisons including full body and upper body motion.
We can see that our framework generates more realistic, natural and smooth motion compared to CaMN\cite{liu2022beat}, MDM \cite{tevet2023human} and DiffuseStyleGesture \cite{yang2023diffusestylegesture}.
The motion generated by CaMN lacks diversity,
the motion generated by MDM is often discontinuous and DiffuseStyleGesture relies on attention leading to unreasonable results.

\noindent \textbf{Style Control.}
We also visualize the results of style control by our framework in the \textit{supplementary video}.
Compared to DiffuseStyleGesture, our framework controls identity and emotion simultaneously and generates more vivid motion.
Furthermore, the style control exhibits clear semantic differences; for example, motion in a happy state are inherently more positive compared to motion in a sad state.

\noindent \textbf{Custom Speech.}
To verify the generalization of our method, we excerpted Martin Luther King's speech “I Have a Dream” \cite{king1986have} and used text-to-speech to obtain the corresponding audio as input. 
The results are shown in the \textit{supplementary video} material.




\bibliographystyle{elsarticle-num}
\bibliography{mybibe}
\end{document}